\documentclass[11pt,a4paper]{article}

\usepackage{a4wide}
\usepackage{amsmath}
\usepackage{amsthm}
\usepackage{amsfonts}
\usepackage{graphicx}
\usepackage{color}
\usepackage{cite}
\usepackage{paralist}

\usepackage{hyperref}
\hypersetup{%
plainpages=false,
colorlinks,urlcolor=blue,linkcolor=blue,citecolor=blue,
pdftitle={Estimation of Joint Distribution of Demand and Available Renewables
       for Generation Adequacy Assessment},
pdfauthor={Chris Dent and Stan Zachary},
pdfsubject={},
pdfpagemode={UseOutlines},
pdfpagelayout={OneColumn},
pdfstartview={FitBH}
}

\theoremstyle{remark}

\newcommand{\elcc}{\nu^{\mathrm{ELCC}}}
\newcommand{\efc}{\nu^{\mathrm{EFC}}}

\renewcommand{\Pr}{\mathbf{P}}
\newcommand{\E}{\mathbf{E}}

\title{Estimation of Joint Distribution of Demand and Available Renewables
       for Generation Adequacy Assessment}

     \author{S.~Zachary\footnote{S. Zachary is with the School of Mathematical and
         Computer Sciences, Heriot-Watt University, Edinburgh EH14
         4AS, UK (Email: s.zachary@hw.ac.uk).}
         \ and C.J.~Dent\footnote{C.J. Dent is with the School of Engineering and
         Computing Sciences, Durham University, DH1 3LE, UK (Email:
         chris.dent@durham.ac.uk).}%
     }

\begin{document}

\maketitle

\begin{abstract}
  In recent years there has been a resurgence of interest in
  generation adequacy risk assessment, due to the need to include
  variable generation renewables within such calculations.  This paper
  will describe new statistical approaches to estimating the joint
  distribution of demand and available VG capacity; this is required
  for the LOLE calculations used in many statutory adequacy studies,
  for example those of GB and PJM.

  The most popular estimation technique in the VG-integration
  literature is `hindcast', in which the historic joint distribution
  of demand and available VG is used as a predictive
  distribution. Through the use of bootstrap statistical analysis,
  this paper will show that due to extreme sparsity of data on times
  of high demand and low VG, hindcast results can suffer from sampling
  uncertainty to the extent that they have little practical meaning.
  An alternative estimation approach, in which a marginal distribution
  of available VG is rescaled according to demand level, is thus
  proposed. This reduces sampling uncertainty at the expense of the
  additional model structure assumption, and further provides a means
  of assessing the sensitivity of model outputs to the VG-demand
  relationship by varying the function of demand by which the marginal
  VG distribution is rescaled.
\end{abstract}

 \emph{Keywords - power system planning; power system
  reliability; risk analysis; wind power}

\maketitle

\section{Introduction}
\label{sec:introduction}

{P}{robabilistic} risk assessment has been used widely
for many years in generation and network capacity planning, both in
industry and in academic research
\cite{banda,anders,endrenyi,libook}. There has been particular focus
recently on the inclusion of variable generation (VG) technologies
within these calculations, on which there is a wide literature on
topics such as wind (\cite{tf09} and references therein), solar
(\cite{solartf} and references therein), tidal \cite{radtke} and wave
\cite{keanewave}.

As with any modelling work, estimation of model inputs from data forms
a critical part of resource adequacy assessment. The literature on
this is vast and cannot be reviewed in full within a single research
paper, however the IEEE Task Force on Capacity Value of Wind Power survey \cite{tf09} and the
recent Special Section of IEEE Transactions on
Power Systems on Adequacy in Power Systems provides a representative
survey \cite{specsect} of current activity. All these papers use model
inputs which must in some way be estimated from data, with
\cite{rob_specsect} placing particular emphasis on time series
modelling of the VG resource.

The approach taken to the statistical modelling of available VG
capacity depends on the modelling outputs which are required. Where
only expected value risk indices are required then, unless significant
storage has a substantial effect of results, it is entirely
sufficient to use a `time-collapsed' or `snapshot' model, which does
not need to consider statistical associations over time in demand and
generation availability. In contrast, where outputs such as frequency
and duration indices (which depend on time correlations in generation
availability and demand) are required, then it is necessary to build a
time series model of VG output, for instance the ARMA model used in
\cite{rob_specsect}.  This paper will consider only the former case,
i.e. time-collapsed models -- practical adequacy studies such as those
in GB \cite{car} and PJM \cite{pjm} typically use this class of model.

Within adequacy models it is often natural to assume
that available conventional capacity is independent of all else, and
it is then necessary to estimate a predictive joint distribution of
demand and available VG relevant to a future point in time. The
most popular approach to this in the VG-integration literature is to use the
empirical historical distribution of demand-net-of-VG \cite{tf09}
(this is sometimes referred to as the `hindcast' approach).  While this
approach by definition accounts for any relationship between demand
and available capacity as seen in the historical data, results are
usually presented without consideration of sampling uncertainty
arising from the finite quantity of data available for statistical
estimation.
Of those papers which do discuss this issue, \cite{tf09} provides
discussion of the need for multiple years of data in order to provide
meaningful model results, but does not discuss in detail quantitative
assessment of uncertainty in model outputs such as risk and capacity
value indices. \cite{haschereq} provides an assessment of how many
years of data are required for a robust calculation in Ireland, but
the approach is limited by an implicit assumption that the longest
available historic record is sufficient to make a precise point
estimate.

Such uncertainty assessment is particularly important in adequacy, as the risk tends to be dominated
by a small number of periods of high demand and low VG availability, and thus
the available quantity of directly relevant data on times of high demand and low VG
is usually small. Without an assessment of this statistical uncertainty,
the relationship between risk model outputs and the real world cannot be assessed.

Bootstrap  analysis of uncertainty \cite{efronbook} places a
confidence interval on a model output such as LOLE or capacity value metrics,
by repeating the output  calculation
using a large number of input datasets of the same size as the
original one, each created by resampling with replacement from within the
original dataset.  
This paper will show how realistic
uncertainty assessments may be obtained
by resampling in blocks of 1 week to account
for serial correlations within the input demand and generation data;
even within the time collapsed model, it is necessary when assessing
confidence intervals on results to account for such serial
correlations in the historical data. This proposed approach is a major advance
on previous work, in that unlike \cite{haschereq} it allows quantitative confidence
intervals to be placed on risk model outputs, and requires no prior assumption as to the quantity of data required.

While the hindcast approach naturally accounts for the demand-VG
relationship as seen in the data, model outputs derived from hindcast
may be driven mostly by a very small number of historical records with
high demand and low VG resource.  Thus, as our analysis demonstrates,
the uncertainty associated with outputs estimated via this approach
may be very high.  This paper proposes an alternative approach, in
which the \emph{marginal} distribution of available VG conditional on
being in the peak season (when extremes of demand might occur) is
estimated, and the distribution of VG output conditional on demand is
constructed by rescaling the marginal VG distribution according to a
factor which depends on the demand level; this scaling function is
estimated from historical data. This provides a degree of statistical
smoothing enabling both considerably more accurate estimation and an
ability to perform sensitivity analysis through varying the parameters
of the scaling function, albeit at the expense of the, relatively
modest, additional model
structure assumptions required.

The paper \cite{DZpmaps} combined a detailed survey of the available
results in the probability theory of capacity values with a brief
preliminary discussion of statistical modelling issues. The present
paper provides a very substantial advance over \cite{DZpmaps} in its
fully detailed treatment of statistical issues, including
consideration of serial associations in data when assessing sampling
uncertainty in model outputs, and possible approaches to remedying the
failures of the hindcast approach at high wind penetrations.

The probability model used for adequacy assessment
is presented in Section~\ref{sec:model}, and then the new statistical
modelling approaches (the fully developed bootstrap analysis, and the
rescaling the VG distribution according to demand) are described in
detail in Section~\ref{sec:stats}.  While these statistical approaches
have recently been used in the GB statutory Electricity Capacity
Assessment Study \cite{car}, this is the first time that the
underlying modelling methodology \emph{or similar approaches} 
has been presented and analysed in
full in the technical literature.  The data used for the examples are
then described in Section~\ref{sect:data}, and the corresponding
results are given in Section~\ref{sect:res}. Finally conclusions and
discussion of wider application are presented in
Section~\ref{sect:conc}.


\section{Mathematical Model}
\label{sec:model}

\subsection{Probability Model}

This paper will consider a \emph{single-area} or spatially-aggregated
model, i.e. it ignores transmission constraints to estimate the
distribution of the surplus of total supply over total demand within a
region, without reference to the geographical locations within the
region of individual supplies and demands.  This is sometimes also
referred to in the power systems literature as a \emph{Hierarchical
  Level 1 (HL1)}, \emph{copper-sheet} or \emph{single bus} model.

The model considers a randomly chosen, or typical, time instant within
the season to be considered, and consists of a specification of the
joint distribution of the random variables $X$, $Y$ and $D$ which
represent respectively \emph{available conventional generation},
\emph{available variable generation} and \emph{demand} at a randomly
chosen point in time. This is referred to as a \emph{time-collapsed}
or \emph{snapshot} model, and is the class of model used in many
practical adequacy studies, e.g. the GB Capacity Assessment Report
\cite{car} explicitly uses the term time-collapsed, and other studies
which output expected value indices and use this class of model are
described in \cite{lolewg,pjm}.

Then the random variable
\begin{displaymath}
  Z = X + Y - D
\end{displaymath}
models the excess of supply over demand at that time.  Statistical
analysis consists of the estimation of the joint distribution of the
random variables $X$, $Y$ and $D$ (see below), and the output of the
model is the estimated distribution of $Z$.

The random variable $X$, modelling available \emph{conventional
  generation}, is assumed to be \emph{independent} of the pair
$(D,\,Y)$.  In a model for a peak season, when there will be limited
planned maintenance due the possibility of very high demand and
associated high prices, this assumption is typically made,
and is natural under most circumstances. While there is
  recent experience of this assumption breaking down (cold weather
  events in Great Britain Feb 2012, Texas Feb 2011, Eastern
  Interconnection winter 2013-14), the quantity of relevant data is so
  minimal that this may only be treated by scenario analysis. In
other seasons this assumption might not be so natural if maintenance
schedules can be flexed in response to anticipated tight margins.  The
marginal distribution of $X$ is taken as an input to the model, i.e.\
we proceed on the basis that it is known; typically the
  distribution of $X$ will be taken as that of the convolution of the
  distributions of the available capacities of the individual
  generating units, those outputs being themselves assumed
  statistically {independent}.

\subsection{Adequacy Indices}

In particular, the \emph{loss of load probability} (LOLP)
and \emph{expected power unserved} (EPU) are defined by
\begin{eqnarray}
  \mathrm{LOLP} & =& \Pr(Z \le 0)\\
  \mathrm{EPU} & =& \E(\max(-Z,\,0) = \int_{-\infty}^0 dz\,\Pr(Z \le z),
\end{eqnarray}
where $\Pr$ and $\E$ denote probability and expectation respectively.
The LOLE is then defined to be LOLP multiplied by the number of demand
periods (e.g. hours or days depending on time resolution of demand
data) in the season modelled, while the EEU is defined to be EPU
multiplied by the length of the season.

If $D$, $X$ and $Y$ are the demand and available generating capacities
at a randomly chosen point in time, this formulation is precisely
equivalent to the more usual specification of LOLE and EEU as
respectively the expected number of periods of shortfall, and the
expected total energy demanded but not supplied, in the future season
under study. The snapshot formulation will prove convenient for
specifying statistical estimation procedures in the next section, and
is used in \cite{DZpmaps,zdjrr} to derive a number of analytical
results associated with generation adequacy and capacity values.

\subsection{Capacity Value Metrics}

It is common also to use capacity value metrics to visualise the
contribution of VG capacity within adequacy assessments. In
general, these specify a deterministic quantity which is equivalent in
adequacy terms to the VG capacity $Y$ within the calculation
under consideration.  

The two most commonly used capacity value metrics are Effective Load
Carrying Capability (ELCC), the additional demand which the VG
resource $Y$ supports while maintaining the same risk level, and
Equivalent Firm Capacity (EFC), the completely reliable capacity which
would give the same risk level if it replaced $Y$.

Defining $M=X-D$, the margin of conventional capacity over demand,
the ELCC $\elcc_Y$ may be defined as the solution of
\begin{equation}
  \Pr(M<0)=\Pr(M+Y<\elcc_Y)
  \label{eq:elccdec}
\end{equation}
and the EFC $\efc_Y$ as the solution of
\begin{equation}
  \Pr(M+Y<0)=\Pr(M+\efc_Y<0).
  \label{eq:efcdec}
\end{equation}
The choice of capacity value metric for a particular application
should be the definition most appropriate to that application, which
may be ELCC, EFC or another alternative.  The metrics ELCC and EFC are
closely related mathematically, and results and techniques applicable
to one may easily be transferred to the other \cite{zdjrr}.

This paper will use EFC for examples of capacity value estimation, as
it visualises the contribution of VG resource~$Y$
within a particular future scenario, and thus
answers a quite generally relevant question when studying such a scenario.
However, in our examples below, similar conclusions would be obtained
were ELCC to be used instead.

\section{Statistical estimation}
\label{sec:stats}


Various assumptions may be made about the statistical
relation between $D$ and $Y$ for the purpose of
estimating the joint distribution of the
pair $(D,\,Y)$. This section will describe the principal options
available, and the consequent assessment of uncertainty in model
outputs. {As discussed in the introduction, without such
uncertainty assessment the relationship between model results and
the real system is unclear, and thus robust interpretation of
modelling results is not possible.}

\subsection{Possible modelling assumptions}

Typically one has available for estimation a time series of historical
observations $(d_t,\,y_t)$ of the pair $(D,Y)$. For demand, an
`observation' will typically mean historical metered demand
appropriately rescaled to the underlying demand levels in the future
scenario under study. For renewables, an `observation' may refer to
either metered historical output rescaled to the scenario under study,
or a synthetic observation of renewable output combining historical
resource (wind speed or solar irradiance) data with the future
installed capacity scenario.


If no assumption is made regarding the statistical relation between
$D$ and $Y$, the distribution of $D-Y$ (i.e.\
\emph{demand-net-of-variable-generation}) is estimated directly from
the \emph{empirical} distribution of the observations $d_t-y_t$ (with
the observations $d_t$ and $y_t$, appropriately rescaled as described
earlier).  The simplest option, where the distribution of $D-Y$ is
taken to \emph{be} the empirical distribution of~$d_t-y_t$, is often
called the \emph{hindcast} approach; it then follows from the
independence of $X$ and all else that, for example, we have
\begin{displaymath}
  \mathrm{LOLP} = \frac{1}{N}\sum_{t=1}^N F_X(d_t - y_t) 
\end{displaymath}
where $F_X$ is the cumulative distribution function of $X$ and $N$ is
the total number of observation periods. The hindcast approach has
been used widely in the literature, for instance by the IEEE Task
Force on Capacity Value of Wind Power \cite{tf09}.



Typically, however, historical measurements corresponding to the
extreme right tail of the distribution of $D-Y$ are extremely
sparse. This means that without making \emph{some} reasonable
assumption about the form of the statistical relationship between $D$
and $Y$, confidence intervals associated with quantities such as LOLE
and EEU (which are determined primarily by this right tail) are so
wide that the results are of limited practical use.

Making some such assumption usually results in considerably narrower
confidence intervals representing sampling uncertainty in model
outputs, as compared to those obtained from the use of hindcast, at the
expense of any additional uncertainty arising from the assumption
itself.  Additionally, a further benefit of assuming a parametrised
form of the $(D,Y)$ relationship may be the ability to perform
transparently an analysis of the sensitivity of model outputs to that
relationship.



The simplest such assumption is that $D$ and $Y$ are
\emph{independent}.
In most systems, a general assumption of year-round independence of
$D$ and $Y$ will be unrealistic for most practical applications;
however, where there is no strong diurnal variability in variable
generation availability (e.g. for wind in the peak winter season in
GB), an assumption of independence conditional on being in the peak
season may be quite reasonable.
  
  
A further possibility is to assume that the distribution of $Y$
\emph{conditional} on~$D$ is given, for each value $d$ of $D$, by the
marginal distribution of $Y$ rescaled by some factor $\lambda(d)$,
i.e.\ by the distribution function $F_{Y|D=d}(y)=F_Y(y/\lambda(d))$
where $F_Y$ is the marginal distribution function of $Y$.  Note that,
given $\lambda(d)$, the marginal density of Y which is being
implicitly assumed in our analysis is given by
\begin{equation}
  \hat f_Y(y) = \int \mathrm{d}d' f_D(d')f_{Y|D=d'}(y)
\end{equation}
(where the pdf $f_{Y|D=d}(y)$ is the derivative of the distribution
function $F_{Y|D=d}(y)$ specified earlier in this paragraph), and this
is not in general quite the density $f_Y(y)$.  However, in the case
where, as here, we take $\lambda(d) = 1$ except for $d$ in a set of
small probability, we have approximate equality between these two
densities.  Further when, again as here, we have $\lambda(d) \le 1$
for all $d$, the approximation is conservative in that it assumes, if
anything, less VG than might be suggested by its empirical marginal
distribution.

When some assumption is made as above about the statistical
relationship between $D$ and $Y$, the marginal distributions of these
random variables are then estimated separately from the available data
for each, and combined with this assumption to obtain the distribution
of $D-Y$. The distribution of $Z$ is then the convolution of the
distributions of $X$ and of $W-D$.

This approach was taken in the 2014 GB Capacity Adequacy Study, where
the reference case (wind-demand independence conditional on being in
the peak season) was contrasted with a sensitivity in which the scale
factor $\lambda(d)$ above was taken to be close to one except for the
most extreme values of demand $d$, where it was  reduced
to as low as $0.5$, corresponding to the assumption of reduced wind at
times of very high demand.

\subsection{Assessment of Uncertainty in Model Outputs}
\label{sec:uncert-assessm}

In the case where the joint distribution of all the variables involved
has an assumed analytical representation -- e.g.\ that these variables
are independent and that the distribution of each has some assumed
parametric form -- we may use standard analytical techniques for the
derivation of standard errors, confidence intervals, etc.  However, it
is usually the case that such an assumed framework is not available.
We may then use instead modern simulation techniques, notably
\emph{bootstrapping} \cite{efronas, efronbook}.  This is applicable to
any adequacy calculation approach, including those described earlier
in this section.

In the case where successive instances of the pair $(D,\,Y)$ may be
treated as independent and identically distributed, i.e.\ there is no
serial association over time, bootstrapping consists of resampling
\emph{with replacement} individual records from the original data;
each such bootstrap sample~$b$ is of the same size~$n$, say, as the
original data set, and for each sample~$b$ the statistic of interest
(e.g.\ the chosen estimate $\efc_Y(b)$ of $\efc_Y$) is
calculated; the distribution of this statistic over a sufficiently
large number of bootstrap samples~$b$ mirrors the sampling properties
of the original estimator, under the assumption that the data are
themselves a set of independent identically distributed observations
(see \cite{efronbook} for a complete description).

However, where there does exist serial association in
the successive observations of the pairs $(D,\,Y)$, as will
typically be the case in practice, then some modification of the
bootstrap procedure is required. One option is to resample from the
original data in \emph{blocks}, where each such block is sufficiently
large that the successive blocks may be treated as at least
approximately independent.



\section{Data for Examples}\label{sect:data}

The examples in this paper are based on National Grid's `Gone Green
2013' (GG) scenario \cite{ukfes} for the GB power system in Winter
2013/14 (National Grid is
the GB Transmission System Operator). As not all fine detail of the GG
scenario can be shared by National Grid, the generating unit
capacities used here are adjusted slightly from those in the original
GG scenario; the scenario used here is thus referred to as `Adjusted
Gone Green' (AGG). It is thus generally representative of GB system
scenarios, and will serve well to illustrate the methodological points
made in this paper.  All other historical data used as inputs to the
calculations are also supplied by National Grid.

\subsection{Demand Data}

Coincident Great Britain wind resource and demand data are available
for the seven winters 2005-12. The demand data are based on the
historical metered demand available publicly at \cite{demanddata}; an
estimate of historical distribution-connected wind output is added on
to each transmission-metered demand record, allowing distribution- and
transmission-connected wind to be treated on a common basis when
studying future scenarios.

Historical demand data may be rescaled to a required underlying level
using each winter's average cold spell (ACS) peak demand, which for
the 2013/14 AGG scenario is 55.55 GW. ACS peak demand is the standard
measure of underlying peak demand level in Great Britain; conditional
on a given underlying demand pattern, it is the median out-turn winter
peak demand \cite{sqss}.

The historical metered demand data are on a half-hourly time
resolution. However as the wind data used here are on a one-hour
resolution, the demand series is converted to hourly observations by
taking for each hour the higher of the two half-hourly demands
contained therein.

In order to account for the operator's practice of taking emergency
measures such as demand reduction in preference to eroding the
frequency response which protects the system against sudden losses of
infeed, 700 MW is added to all demand values to balance the capacity
of conventional generation that is used to provide this response.
This is consistent with the statutory Capacity Assessment Study
\cite{car}.


\subsection{Wind Data}

The GB aggregate wind power historical time series
combines wind speed resource data
from NASA's MERRA reanalysis dataset with the AGG
scenario for installed wind capacity in 2013/14
(in which aggregate installed capacity is 10120 MW).
These data may thus be treated as observations of historical
available wind capacity assuming this capacity scenario.

For the examples presented, these wind power data are rescaled to
different installed capacities while maintaining the same time series
of wind load factor. This permits study of the dependence of results
on total installed wind capacity while holding other data constant.
The origin of this dataset is described in more detail in \cite{car}.



\subsection{Conventional Plant Data}

The available capacity from each individual unit is assumed to follow
a two-state distribution with either zero or maximum capacity
available. The unit capacities are taken from the 2013 AGG scenario,
and the availability probability for each class of unit is the central
estimate from the Great Britain Electricity Capacity Assessment Report
\cite{car}.
 
The distribution of available conventional capacity is then
constructed assuming that the availabilities of the different units
are statistically independent.  The result of convolving these
distributions is usually referred to in power systems as a capacity
outage probability table \cite{banda}. For the 2013/14 scenario used,
the distribution of available conventional capacity $X$ has mean 58820
MW and standard deviation 1950 MW.

As the focus of this paper is on the treatment of wind and demand
data, the distribution of $X$ will not be discussed further here.
Examples of sensitivity of LOLE outputs to the distribution of $X$ may
be found in \cite{car}, and further discussion of statistical issues
associated with constructing the distribution of $X$ may be found in
\cite{nonit}.

\subsection{Data Visualisation}\label{sect:vis}
\begin{figure}%
\begin{center}
\includegraphics[width=10cm]{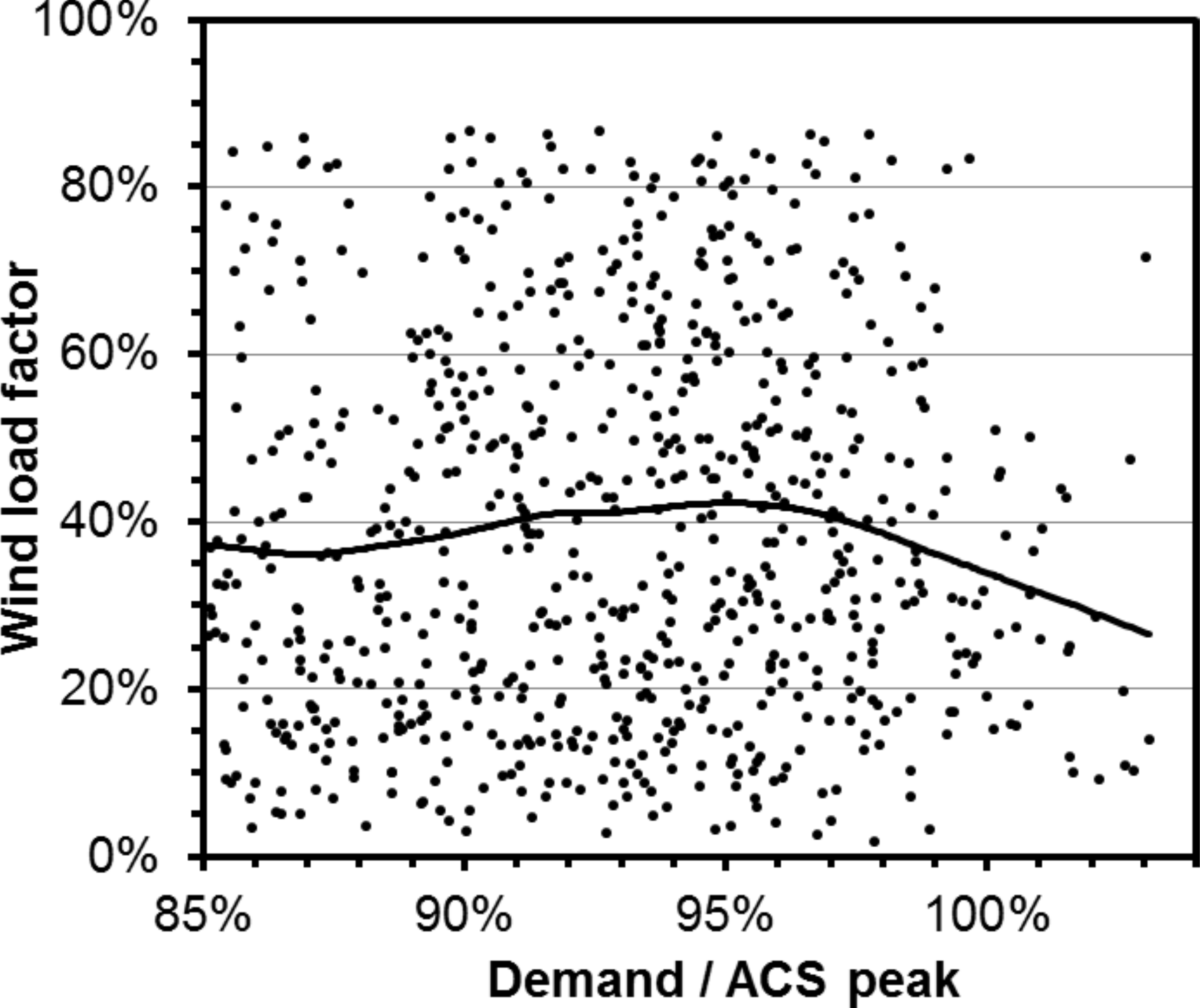}%
\caption{Wind load factor against demand at time of daily peak in
  winters 2005-12, and LOESS smoothed curve.}%
\label{fig:stats_scatter}%
\end{center}
\end{figure}
The available demand and wind data from times of high demand are
visualised in Fig~\ref{fig:stats_scatter}.  The plot shows system-wide
available wind load factor against demand (each demand data point is
normalised by that winter's ACS peak, and we consider those data
points for which the normalised demand exceeds 90\%).  For clarity,
only data points from times of daily peak are plotted.
The smooth solid line is an estimate of the mean wind
load factor conditional on the daily peak demand. This is fitted using the
LOESS local regression method \cite{loess},
and some experimentation shows that 
it is robust against varying choices of the width of the smoothing
window.

It is clear from this plot that the quantity of data arising from
times of truly extreme demand is very small.  Indeed, of the 30 days
on which ACS peak was exceeded during these seven winters, 23 were in
January 2009 and December 2010 and a further 5 in December 2008--January
2009. Due to the serial associations inherent in weather patterns,
there is only one completely independent data point provided by
each of these periods in 2009 and 2010.


\section{Results on Statistical Estimation and Uncertainty Analysis}\label{sect:res}

\subsection{Bootstrap  Analysis of Uncertainty for LOLE and EFC Results}

\subsubsection{Bootstrap analysis}

This section provides an example application of the parameter
estimation and bootstrap uncertainty assessment procedures described
in Section~\ref{sec:stats}.  The wind and demand data, and the
conventional plant distribution, are as described in
Section~\ref{sect:data}.  Estimation both of the LOLE and wind EFC (as
defined in Section \ref{sec:model}) is considered.

The calculations consider only the peak season, which in GB dominates
the adequacy risk; a similar bootstrap approach could be applied to
wind and demand data in other seasons, in the context of a year-round
calculation which models maintenance explicitly. From each of the
seven winters, 20 weeks of coincident wind and demand data are used,
beginning on the first Sunday in November. In each winter a 2-week
Christmas block is defined, giving in the data set as a whole seven
2-week Christmas blocks and 126 1-week `normal' blocks, each block
running from Sunday to Saturday.

For each of the statistical approaches described above, once parameter
estimates are obtained (e.g. of EFC or LOLE), bootstrap resampling may
be used to make to 
corresponding assessments of uncertainty, as described in
Section~\ref{sec:stats}.  In each case we use resampling from the
original data to generate 1000 bootstrap samples of the same
size as that of the original 7-year data set; for each such sample we
may readily calculate model outputs such as EFC, LOLE, etc. As noted
previously, the variability of these 
estimates over the bootstrap samples provides the corresponding
uncertainty assessments associated with the original central
estimates of the model outputs.  

For the hindcast approach, each bootstrap sample is obtained by
resampling (with replacement) seven 2-week Christmas blocks and 126
1-week `normal' blocks from the bivariate time series of historical
wind and demand.  Using the alternative approaches of assuming
wind-demand independence or rescaling the marginal distribution of
wind according to demand level, each bootstrap sample is obtained by
sampling (again with replacement) independently from the metered wind
and demand time series. For demand, 126 `normal' 1-week blocks and 7
Christmas 2-week blocks are resampled; however for wind, 140 week-long
blocks are sampled from the 140 week long blocks in the full time
series (unlike demand, wind data from the Christmas period has no
special status).

\subsubsection{Results}

\begin{figure}%
\begin{center}
  \includegraphics[width=10cm]{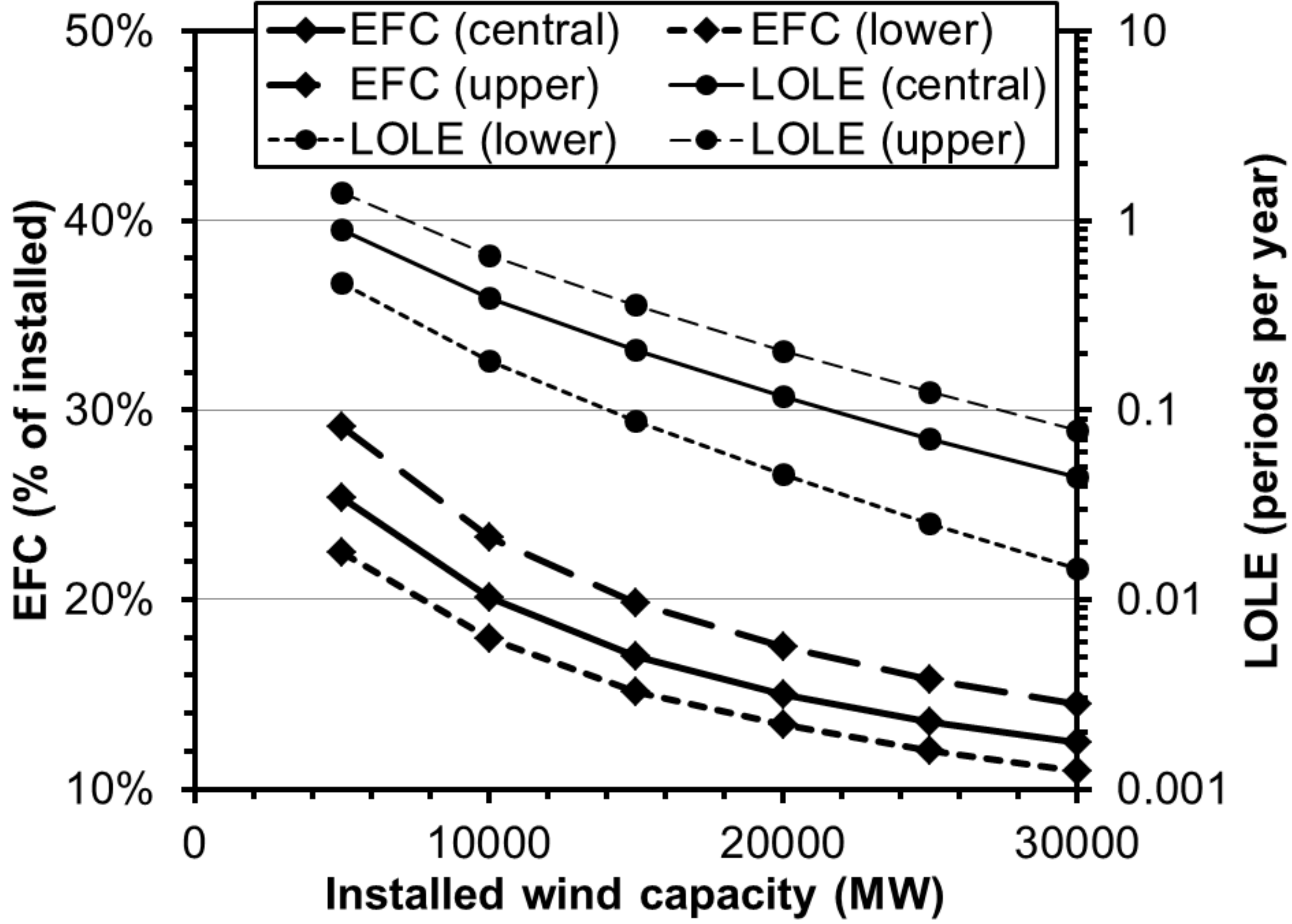}%
  \caption{LOLE and EFC results based on hindcast, along with 95\%
    confidence intervals derived from bootstrap resampling.}%
  \label{fig:bs_hcast}%
\end{center}
\end{figure}

\begin{figure}%
\begin{center}
  \includegraphics[width=10cm]{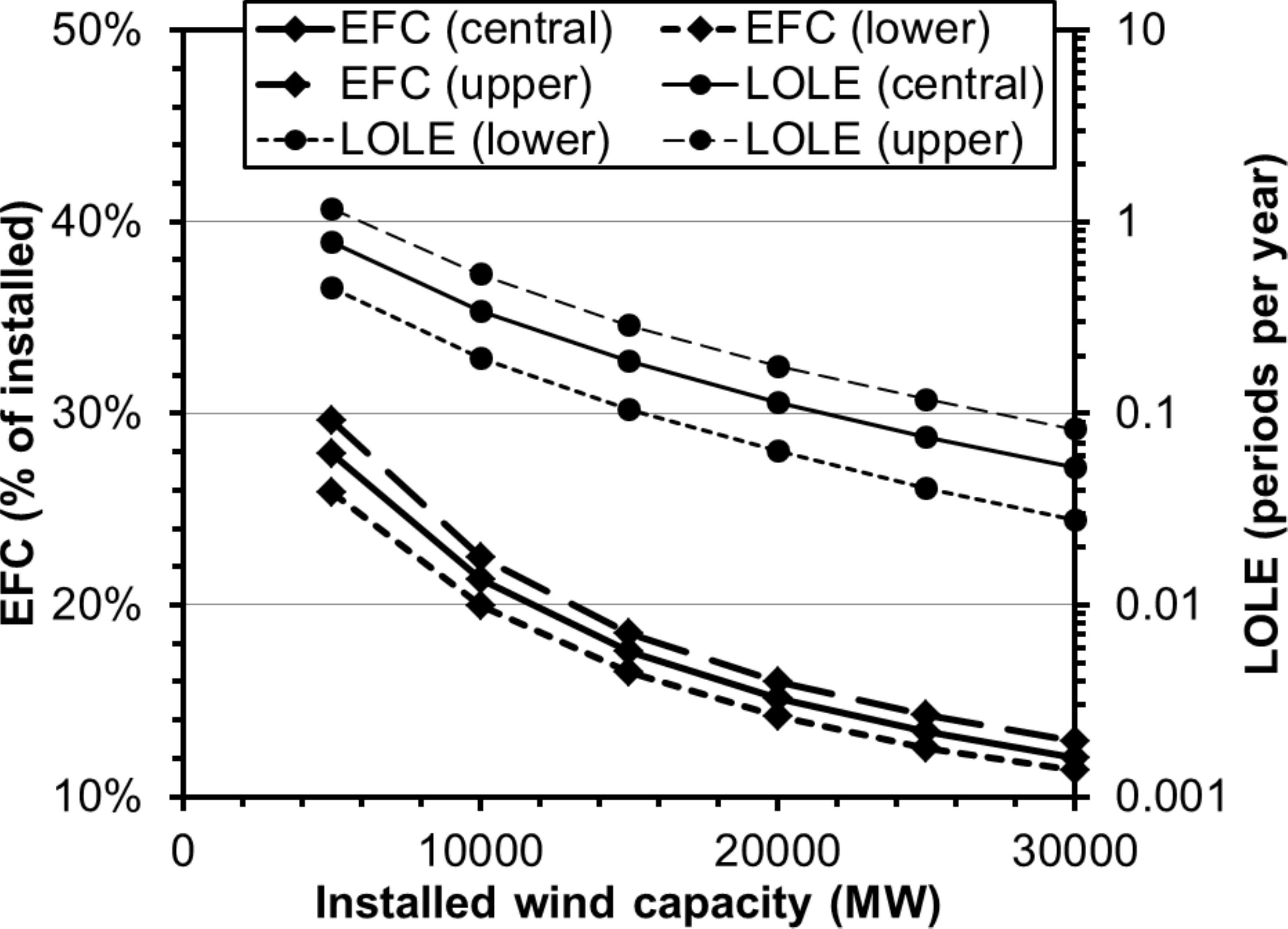}%
  \caption{LOLE and EFC results based on assuming wind-demand
    independence, along with 95\% confidence intervals derived from
    bootstrap resampling.}%
  \label{fig:bs_ind}%
\end{center}
\end{figure}

Figures~\ref{fig:bs_hcast} and \ref{fig:bs_ind} show LOLE and EFC
results derived respectively from hindcast estimation and using the
assumption of wind-demand independence, in each case along with 95\%
confidence intervals derived via bootstrap resampling.
EFC is shown as a percentage of installed capacity.  

\begin{figure}%
\begin{center}
  \includegraphics[width=10cm]{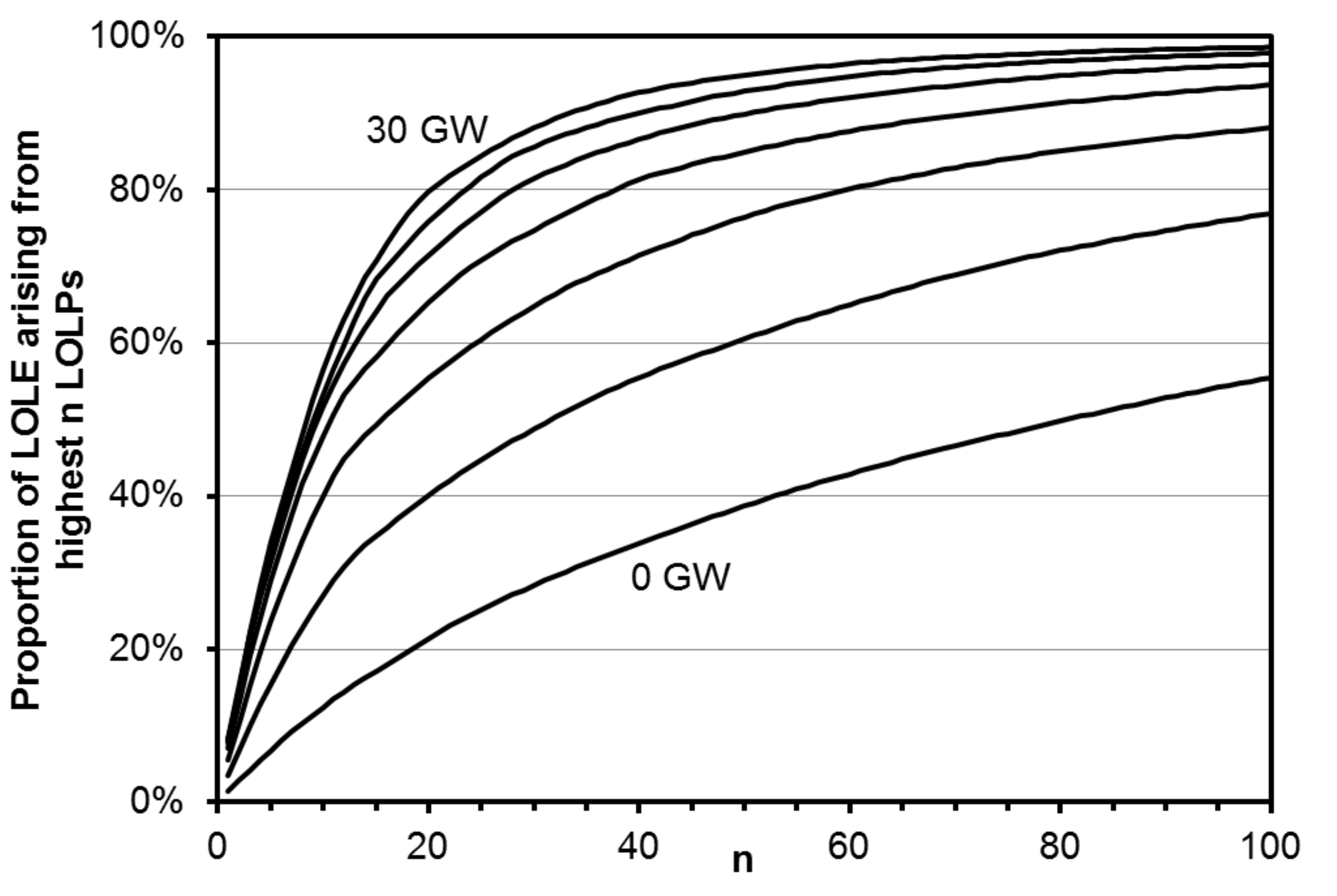}%
  \caption{Proportion of hindcast LOLE arising from the $n$ historical hours with
  the highest net demand $d_t-y_t$ (or equivalently the highest LOLPs). The 7 series are
  for installed wind capacities from 0 GW to 30 GW in steps of 5 GW.}%
  \label{fig:stats_topn}%
\end{center}
\end{figure}

Features of particular note in the plots are:
\begin{itemize}
\item In the hindcast calculation, the ratio of the upper and lower
  ends of the LOLE confidence interval increases very substantially as
  the installed wind capacity increases, until at 30 GW installed
  capacity the ratio between the CI's upper and lower limits is
  4.6. This is a consequence of how, in a hindcast calculation,
  only the very small number of historical records with both high demand
  and low wind load factor make a substantial contribution.
\item In the calculation which assumes wind-demand independence, the
  ratio of the upper and lower limits of the LOLE CI barely changes as
  the wind capacity increases.  In this case, while data on times of
  very high demand remain sparse,
  the quantity of wind data used in estimating the distribution of $Y$
  is comparatively large.  The consequences of data sparsity within
  the EFC and LOLE calculations hence increase only very slightly as
  the installed wind capacity increases.
\item The relative uncertainty in EFC appears much less than that in
   LOLE. There is some degree of cancellation between the uncertainties
   in the distributions of demand on each side of the equation
   defining EFC. Due to the highly nonlinear dependence of LOLE
   on shifts of the distribution of margin, there is also a tendency for
   precisely the same uncertainty to appear less substantial when expressed
   in MW terms through capacity values. 
\end{itemize}
As discussed above, the hindcast results suffer at large wind
capacities from such large sampling uncertainty that they are of very
little value.  The origin of this is explored further in
Fig.~\ref{fig:stats_topn}, in which it is seen that at large installed
wind capacities in this GB example LOLE results are driven by a very
small number of items of historical data combining high demand with
low wind resource.  Indeed at 30 GW installed wind capacity 2/3 of the
calculated LOLE arises from data from just 6 days (1 in each of winters 06-07, 09-10 and 11-12,
and 3 in 10-11).


While sampling uncertainty is much reduced in the `wind-demand
independence' results, that assumption introduces additional
uncertainty, arising from the possibility that there might indeed be
some substantial wind-demand statistical relationship which cannot be
quantified with the available data. 
This problem is addressed in the next section.

\subsection{Rescaling of Wind Distribution}

\begin{figure}%
\begin{center}
  \includegraphics[width=10cm]{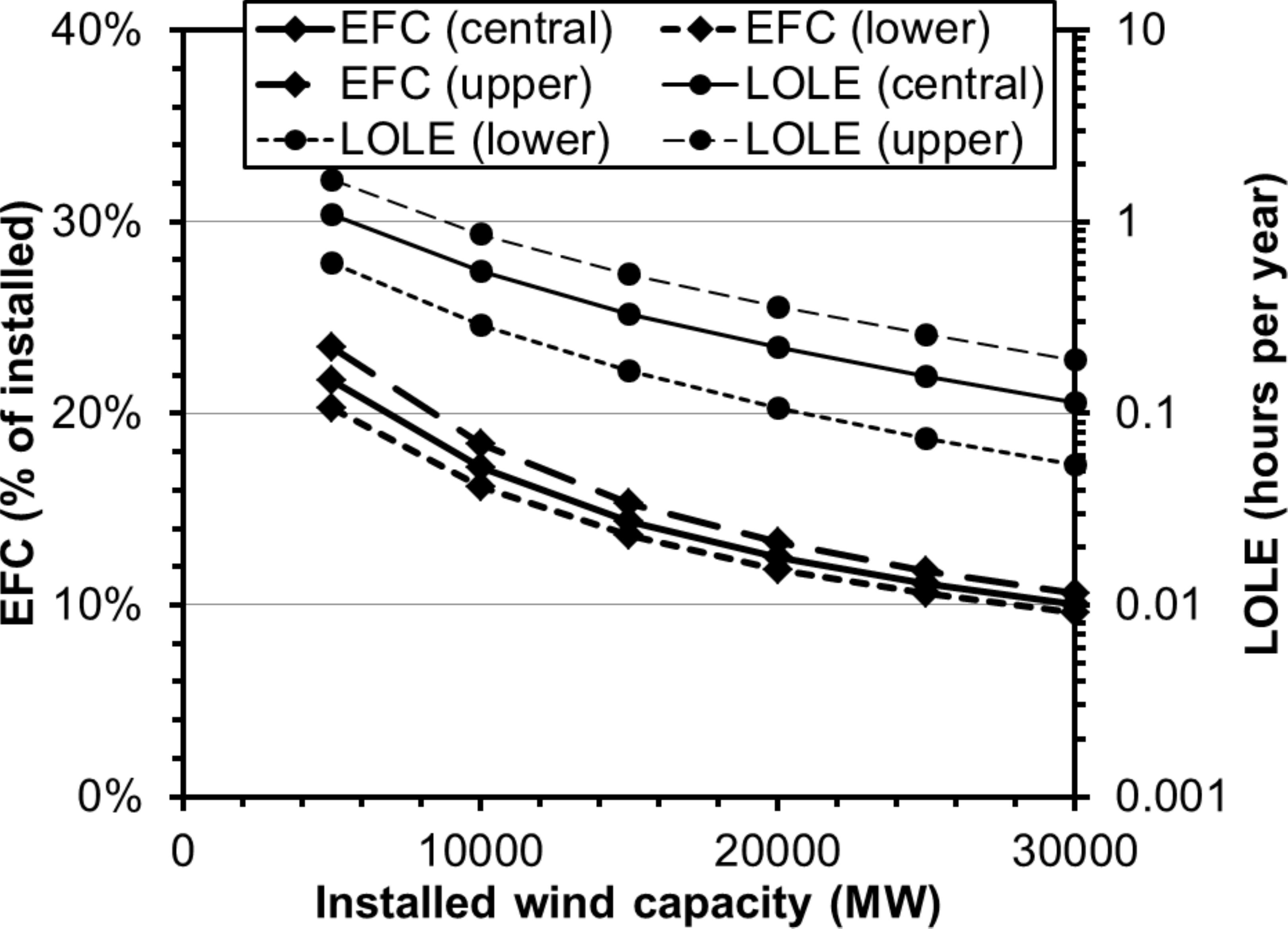}%
  \caption{LOLE and EFC results based on rescaling distribution
  of available wind according to demand level, with
  a scaling factor of 0.5 at the highest demands.}%
  \label{fig:bs_scale}%
\end{center}
\end{figure}

If the distribution of $Y$ is to be rescaled by a demand-dependent
factor $\lambda(d)$ as described in Section~\ref{sec:stats}, then
$\lambda(d)$ may, for example, be estimated based on a wind v.\ demand
plot and smoothed relationship as in Fig.~\ref{fig:stats_scatter}.
For the available data, this figure shows a decrease from a mean load
factor of around 40\% on days of relatively modest demand, to a level
of 20--30\% at the times of very highest demand.  However, as
discussed in Section \ref{sect:vis}, the data corresponding to the
very high demands almost all arise from just three distinct periods in
the three winters 2008-11.  The data from these winters cannot be said
to provide more than three independent data points, and are thus
insufficient to provide a point estimate of $\lambda(d)$ with any real
confidence.

We can however estimate a range of credible model outputs (i.e.\ LOLEs
and EFCs) by providing credible optimistic and pessimistic bounds on
the scaling function $\lambda(d)$ corresponding to the highest
demands. As there is no evidence to suggest that the wind resource is
better at times of very high demand compared to that across the peak season
as a whole, a credible upper bound is $\lambda(d)=1$, i.e.\
wind-demand independence. We take as a credible lower bound a
rescaling function~$\lambda(d)$ of the form
\begin{equation}
  \label{eq:6}
  \lambda(d) =
  \begin{cases}
    l_1, & \qquad d \le d_1,\\
    l_1 + \dfrac{d-d'_1}{d'_2-d'_1}(l_2-l_1), & \qquad d'_1 < d < d'_2,\\
    l_2, & \qquad d \ge d'_2
  \end{cases}
\end{equation}
where $d'_1=0.95$, $d'_2=1.03$, $l_1=1$, and $l_2=0.5$.  This is
substantially more pessimistic than that suggested by the smoothed
relationship in Fig.~\ref{fig:stats_scatter}; we do not believe that
the data provide a justification for considering a relationship which
is more pessimistic still.

Results for this pessimistic bound on wind's contribution are displayed in
Fig.~\ref{fig:bs_scale}.
In this case the extent of the confidence
intervals increase very slightly as the installed wind capacity increases,
and the `rescaled $Y$' approach has thus achieved its aim (as compared to hindcast)
of giving much reduced sampling uncertainty in exchange for an additional
natural model structure assumption.

Performing this pessimistic rescaling of the distribution of available wind
increases the central estimate of
LOLE by a factor which ranges from 1.4 (relative to the wind-demand-independence
case) at 5 GW of
installed wind capacity, to 2.2 at 30 GW installed capacity.  However,
the interpretation of the EFC plot is more transparent. The difference
between the EFC of the wind with and without the pessimistic rescaling
is 6.2\% of installed capacity (310 MW) for 5 GW installed wind
capacity, and 2.0\% of installed capacity (600 MW) at 30 GW installed
wind.

It should be remembered that the choice of these optimistic and
pessimistic bounds on the scaling function $\lambda(d)$ arise from
very considerable uncertainty (i.e.\ very sparse data indeed)
regarding the wind resource at the times of the very highest demands.
However, the corresponding variation of the EFC result (just $\pm 300$
MW at 30 GW installed capacity) demonstrates that compared to other
uncertainties in inputs the consequences of this uncertainty in the
wind-demand relationship are relatively unimportant; examples of such
more important uncertainties include
\begin{itemize}
\item uncertainty in future ACS peak level is of order 1 GW or more;
\item uncertainty over which conventional units will commission or
  decommission -- to leading order, adding or withdrawing a single
  unit or station shifts the distribution of available capacity by its
  rated capacity, which may be up to around 1 GW in the case of CCGT
  or 1.8 GW in the case of nuclear;
\item uncertainty in conventional unit availability probabilities --
  to leading order, a 1 percentage point change in the mean available
  conventional capacity shifts the distribution of available capacity
  by around 700 MW.
\end{itemize}
These are all substantially larger than the range of credible wind EFCs
arising from uncertainty in the wind-demand relationship.

\section{Conclusions}
\label{sect:conc}

This paper has described new statistical approaches to the estimation
of inputs and assessment of uncertainty in outputs of generation
adequacy risk models.  Bootstrap analysis provides a generally
applicable statistical means of quantifying the consequences of
sampling uncertainty; a key observation is that where data on the
coincidence of high demand and low VG are very sparse, sampling
uncertainty in hindcast calculations may be so great as to render
model results quite meaningless.  A new approach of rescaling the
marginal distribution of available VG according to demand level is
thus proposed; this adds an additional -- but not unreasonable -- model
structure assumption, but considerably reduces sampling uncertainty;
the use of alternative scaling functions further facilitates analysis
of the sensitivity of model outputs to variation in the assumed
wind-demand relationship.

In using this bootstrap analysis, one must remember that it only
assesses one aspect of uncertainty in these calculations, namely that
arising from the use of a finite-size data set within a given
statistical estimation process. Other uncertainties which should be
considered include that in the distribution of available conventional
capacity, how representative historical demand data are of future
underlying demand patterns, if synthetic wind data are used how well
they represent real wind farm outputs, and possible consequences of
natural climate variability and human-induced climate change.

{This paper has presented examples for the GB system only,
however the methods presented here are generally applicable.
The bootstrap approach could be used directly in any system;
the rescaling of the VG distribution according to demand level
can also be used in any system.
Some additional care may need to be taken when, for example, 
there is a substantial diurnal variability in the VG resource, since demand already 
has such variation.  One possibility is to condition appropriately on the time of day.

It would be interesting to investigate
how the statistical analysis and degree of sampling uncertainty might
differ between small and large systems, between summer- and
winter-peaking systems (in particular there tends to be stronger
diurnal variability of the wind resource in summer-peaking systems),
and in systems in which solar generation makes a significant
contribution.}

{To conclude, the authors strongly recommend that adequacy studies should
include an assessment of statistical uncertainty in model outputs, for instance using
the methods in this paper. As emphasised above, 
without such
uncertainty assessment the relationship between model results and
the real system is unclear, and thus robust interpretation of
modelling results is not possible.}

\section*{Acknowledgements}

The authors express their thanks to colleagues at National Grid and
Ofgem for many valuable discussions during the GB Electricity Capacity
Assessment project, and express particular thanks to M.~Roberts for
providing the Adjusted Gone Green Scenario data.  They also
acknowledge discussions with C.~Gibson, the IEEE LOLE Working Group,
and colleagues at Durham and Heriot-Watt Universities, EPRI, NREL and
University College Dublin.

This work was supported by National Grid, and by the following EPSRC grants:
EP/I017054/1, EP/E04011X/1, EP/H500340/1 and EP/I016953/1.



\bibliography{jrr}{}
\bibliographystyle{IEEEtran}

\end{document}